\documentclass[prl,twocolumn,showpacs,floatfix,amsmath,amssymb,superscriptaddress]{revtex4-1}
\usepackage{amsfonts}
\usepackage{stmaryrd}
\usepackage{bbm}
\usepackage{mathrsfs}
\usepackage{tipa}
\usepackage{amssymb}
\usepackage{txfonts}
\usepackage{graphicx}
\usepackage{dcolumn}
\usepackage{epstopdf}
\usepackage[colorlinks,linkcolor=blue,urlcolor=blue,citecolor=blue]{hyperref}
\usepackage{multirow}
\usepackage{subfigure}
\usepackage{url}

\begin{document}
\newcommand*{\cm}{cm$^{-1}$\,}
\newcommand*{\TC}{T$_c$\,}
\newcommand*{\TCDW}{T$_{CDW}$\,}
\newcommand*{\A}{2$\Delta$/k$_{B}$T$_{CDW}$\,}
\newcommand*{\Sr}{Sr$_{3}$Rh$_{4}$Sn$_{13}$\,}
\newcommand*{\Ca}{(Sr$_{0.5}$Ca$_{0.5}$)$_{3}$Rh$_{4}$Sn$_{13}$\,}
\newcommand*{\SrCa}{(Sr$_{x}$Ca$_{1-x}$)$_{3}$Rh$_{4}$Sn$_{13}$\,}

\title{Optical spectroscopy study of charge density wave order in Sr$_{3}$Rh$_{4}$Sn$_{13}$ and (Sr$_{0.5}$Ca$_{0.5}$)$_{3}$Rh$_{4}$Sn$_{13}$}
\author{W. J. Ban}
\affiliation{Institute of Physics, Chinese Academy of Sciences, Beijing 100190}

\author{H. P. Wang}
\affiliation{Institute of Physics, Chinese Academy of Sciences, Beijing 100190}

\author{C. W. Tseng}
\affiliation{Department of Physics, National Cheng Kung University, Tainan 70101}

\author{C. N. Kuo}
\affiliation{Department of Physics, National Cheng Kung University, Tainan 70101}

\author{C. S. Lue}
\affiliation{Department of Physics, National Cheng Kung University, Tainan 70101}

\author{N. L. Wang}
\email{nlwang@pku.edu.cn}
\affiliation{International Center for Quantum Materials, School of Physics, Peking University, Beijing 100871}
\affiliation{Collaborative Innovation Center of Quantum Matter, Beijing}

\begin{abstract}
We perform optical spectroscopy measurement on single-crystal samples of Sr$_{3}$Rh$_{4}$Sn$_{13}$ and (Sr$_{0.5}$Ca$_{0.5}$)$_{3}$Rh$_{4}$Sn$_{13}$. Formation of CDW energy gap was clearly observed for both single-crystal samples when they undergo the phase transitions. The existence of residual Drude components in $\sigma_1(\omega)$ below \TCDW indicates that the Fermi surface is only partially gapped in the CDW state. The obtained value of 2$\Delta$/K$_{B}$T$_{CDW}$ is roughly 13 for both Sr$_{3}$Rh$_{4}$Sn$_{13}$ and (Sr$_{0.5}$Ca$_{0.5}$)$_{3}$Rh$_{4}$Sn$_{13}$ compounds, which is considerably larger than the mean-field value based on the weak-coupling BCS theory. The measurements provide optical evidence for the strong coupling characteristics of the CDW phase transition.
\end{abstract}


\maketitle
\section{introduction}

Collective quantum phenomena, such as charge density waves (CDW¡¯s) and spin density waves, are among the most fascinating phenomena in solids and have been a subject of considerable interest in condensed matter physics. The CDWs are mediated by the electron-phonon coupling and are predominantly driven by the nesting of Fermi surfaces (FSs), that is, the presence of two patches of almost parallel FSs that are separated by a wave vector $\textbf{q}_{CDW}$. A single particle energy gap would form in the nested region of the FSs in the CDW ordered state. CDW instability usually appears in low dimensional materials, for example, in quasi one-dimensional (1D) compounds, \emph{e.g.} K$_{0.3}$MoO$_{3}$ \cite{PhysRevB.30.1971}, and 2D transition metal dichalcogenides\cite{ISI:A1989T890400023}, on which the FSs could be easily nested. Upon increasing the dimensions of the samples, the FSs usually become more complex and the nesting conditions become less favored.

Although it is hard to form CDW in 3D materials, recent investigations on A$_{3}$T$_{4}$Sn$_{13}$ (where A=La, Sr, Ca and T=Rh and Ir) with cubic Yb$_{3}$Rh$_{4}$Sn$_{13}$-type structure revealed interesting coexistence of structural phase transition and superconductivity \cite{1367-2630-17-3-033005,PhysRevB.92.024101,PhysRevB.92.195122,PhysRevB.93.134505,PhysRevB.93.245119,PhysRevB.93.245126,1367-2630-18-7-073045,doi:10.1143/JPSJ.79.113705,PhysRevB.83.184509,doi:10.1143/JPSJS.80SA.SA114,PhysRevB.86.024522,PhysRevB.86.064504,PhysRevB.88.104505,PhysRevB.88.115113,PhysRevB.89.125111,PhysRevLett.115.207003,PhysRevB.89.075117,PhysRevB.90.144505,PhysRevB.90.035115,PhysRevB.89.094520,PhysRevLett.114.097002,PhysRevLett.109.237008,PhysRevB.93.245119}. The structural phase transition leads to the formation of a superlattice modulation, which has a lattice parameter twice of that of the high temperature phase. It has been further argued that this superlattice transition is associated with a CDW transition of the conduction electron system. For Sr$_{3}$Ir$_{4}$Sn$_{13}$ and Sr$_{3}$Rh$_{4}$Sn$_{13}$, systematic variation of superconducting transition temperature \TC and CDW phase transition temperature \TCDW have been observed when the unit cell volume of the crystal is varied via chemical substitution. When Sr is partially replaced by Ca, which simulates a positive chemical pressure on the crystals, \TCDW is suppressed whereas \TC increases \cite{PhysRevLett.114.097002,PhysRevLett.109.237008,PhysRevLett.115.207003,PhysRevB.89.075117}.

Optical spectroscopy is a powerful bulk-sensitive technique to detect the energy gaps in ordered state. It yields a great wealth of information about charge excitations and dynamical properties in CDW systems. The recent optical spectroscopy study on Sr$_{3}$Ir$_{4}$Sn$_{13}$ with \TC = 5 K and \TCDW =147 K revealed an energy gap-like suppression in optical conductivity below the structural phase transition temperature. The extracted value of 2$\Delta$/k$_{B}$T$_{CDW}$ is significantly higher than the BCS mean-field value of 3.5 for a density-wave phase transition, pointing towards a strong coupling charge density wave order\cite{PhysRevB.90.035115}. With the purpose to know whether the unconventional strong coupling CDW phase transition commonly exists in such intermetallic compounds, we performed optical spectroscopy measurements on Sr$_{3}$Rh$_{4}$Sn$_{13}$ and (Sr$_{0.5}$Ca$_{0.5}$)$_{3}$Rh$_{4}$Sn$_{13}$, the sister compound of Sr$_{3}$Ir$_{4}$Sn$_{13}$. The two compounds have the superconducting transition temperatures \TC $\approx$ 4.2 K and 7K and CDW transition temperatures \TCDW $\approx$ 138K and 55K, respectively \cite{PhysRevLett.115.207003,PhysRevLett.114.097002,PhysRevB.91.165141}. Our measurement revealed a depletion of low frequency spectral weight due to the formation of CDW energy gap with a significantly large ratio of 2$\Delta$/k$_{B}$T$_{CDW}\sim 13$, yielding optical evidence for a strong coupling CDW order.

\section{EXPERIMENTS}

The Sr$_{3}$Rh$_{4}$Sn$_{13}$ and (Sr$_{0.5}$Ca$_{0.5}$)$_{3}$Rh$_{4}$Sn$_{13}$ single crystals were grown by Sn self-flux method. Detailed descriptions about crystal growth and characterizations were presented in Ref. \cite{PhysRevB.91.165141}. The optical reflectance measurements were performed on as-grown shinny surfaces of the single crystals with a combination of Bruker 113v, Vertex 80v and grating-type spectrometers in the frequency range from 40 to 40000 \cm. An \emph{in-situ} gold and aluminum over-coating technique was used to get the reflectance \cite{Homes:93}. The measured reflectance was then corrected by multiplying the available curves of gold and aluminum reflectivity at different temperatures. The real part of conductivity $\sigma_1(\omega)$ was obtained by the Kramers-Kronig transformation of R($\omega$) employing an extrapolation method with X-ray atomic scattering functions \cite{PhysRevB.91.035123}. This approach of Kramers-Kronig transformation was proved to be more accurate and effective in the analysis.

\section{RESULTS AND DISCUSSIONS}

\begin{figure}[htbp]
  \centering
  \includegraphics[width=7.5cm]{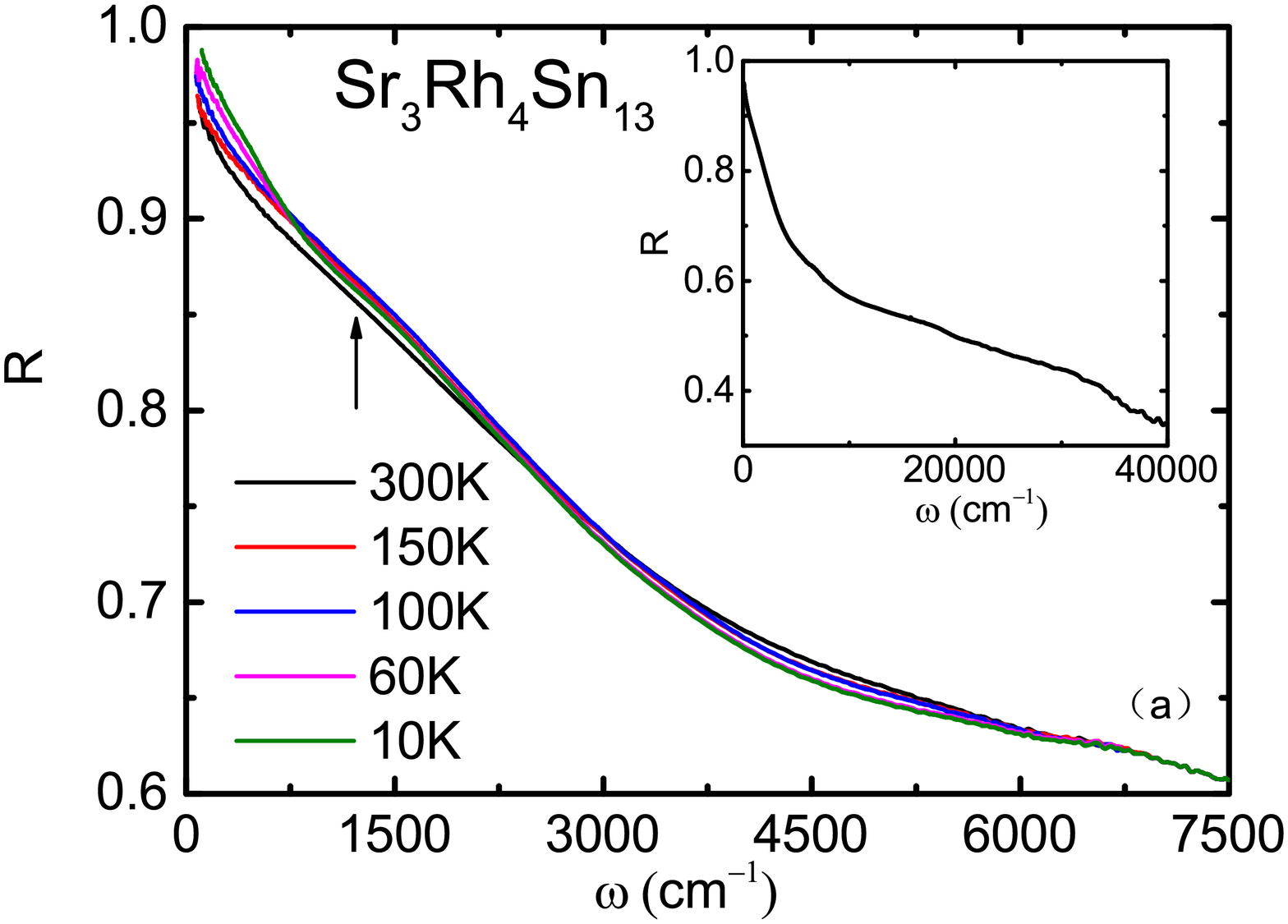}\\
  \includegraphics[width=7.5cm]{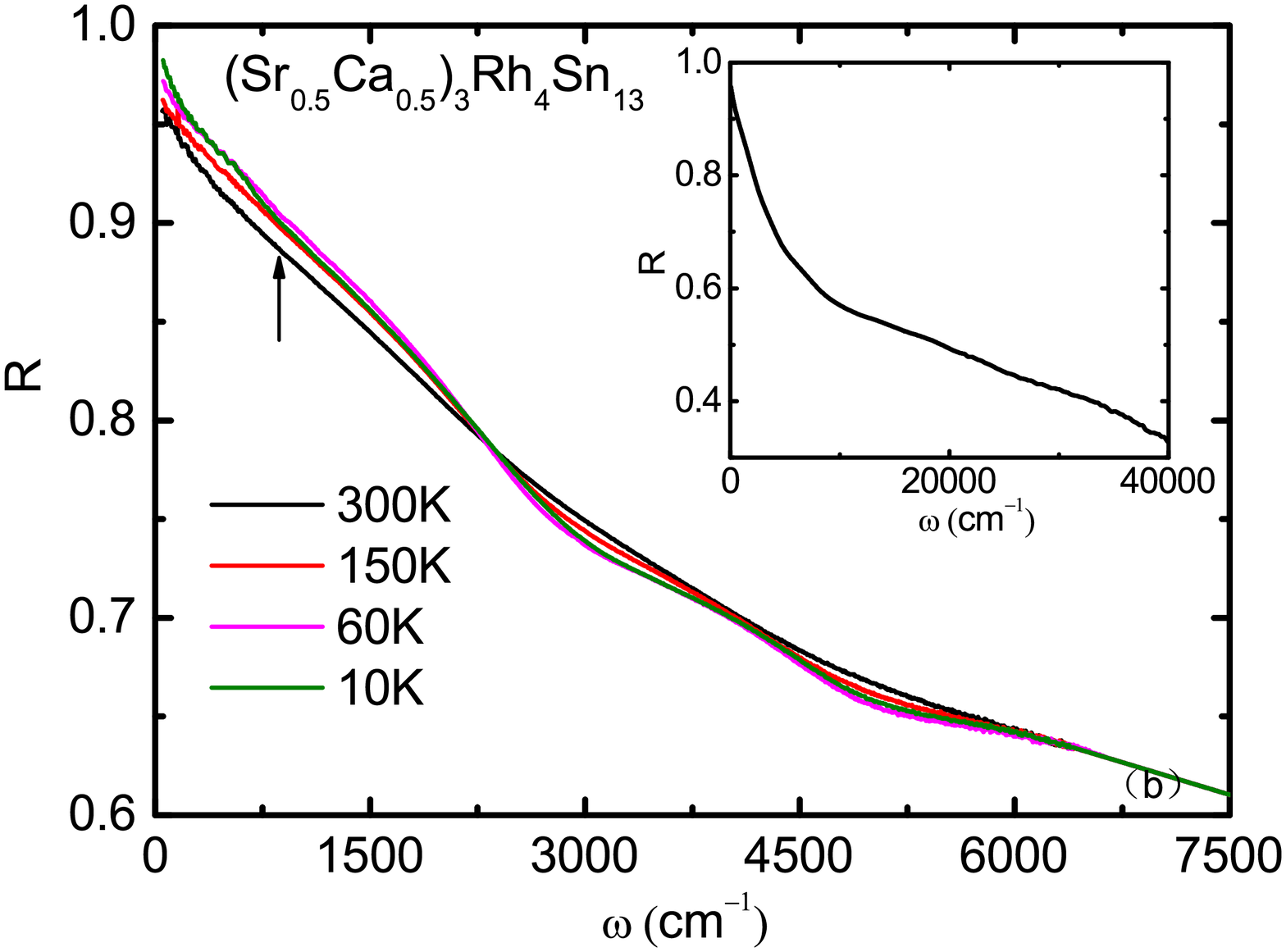}\\
    \caption{The temperature dependent reflectance of Sr$_{3}$Rh$_{4}$Sn$_{13}$ (a) and (Sr$_{0.5}$Ca$_{0.5}$)$_{3}$Rh$_{4}$Sn$_{13}$ (b). The inset shows R($\omega$)
data up to 40 000 \cm at 300 K. The arrows mark the region where the reflectance appears to be suppressed below \TCDW phase transitions.}\label{Fig:ref}
\end{figure}

Figure \ref{Fig:ref} shows the reflectance spectra R($\omega$) of Sr$_{3}$Rh$_{4}$Sn$_{13}$ and (Sr$_{0.5}$Ca$_{0.5}$)$_{3}$Rh$_{4}$Sn$_{13}$ over a broad energy scale at various temperatures. Because gold (or aluminum) depositions were performed only at room temperature, R($\omega$) at room temperature, being obtained from the ratio of measurements just before and after the deposition, has the highest data quality and accuracy. For the low temperature measurement, the sample (and also the reference) have to be cooled down twice in order to obtain the reflectance spectra. As a consequence the data have relatively higher uncertainty and noise level, in particular at very low frequencies. Therefore we need to choose a suitable cutoff frequency for the low frequency extrapolation in Kramers-Kronig transformation. As seen from Fig. \ref{Fig:ref}, the value of R($\omega$) approach unity at zero frequency, both above and below the \TCDW phase transition, indicating they are both metallic in nature. By lowering the temperature, we do not see any sharp change in the optical spectra, but the difference between them is evident. For Sr$_{3}$Rh$_{4}$Sn$_{13}$, the low frequency R($\omega$), roughly below 800 \cm, shows an upturn below the \TCDW phase transition, while the R($\omega$) values between 1000 and 1300 \cm (marked by an arrow) are somewhat suppressed. Those features are resolved more clearly for the spectrum at 10 K. The spectral changes signal the opening of a partial energy gap below the \TCDW phase transition, as we shall see more clearly in the optical conductivity spectra below. Similar spectral evolutions are also seen in (Sr$_{0.5}$Ca$_{0.5}$)$_{3}$Rh$_{4}$Sn$_{13}$ sample although the suppression is weaker and locates at lower frequencies. The spectral features at higher energy scales, for example near 3000 \cm and 5000 \cm, should come from interband transitions.

\begin{figure}[htbp]
  \centering
  \includegraphics[width=7.5cm]{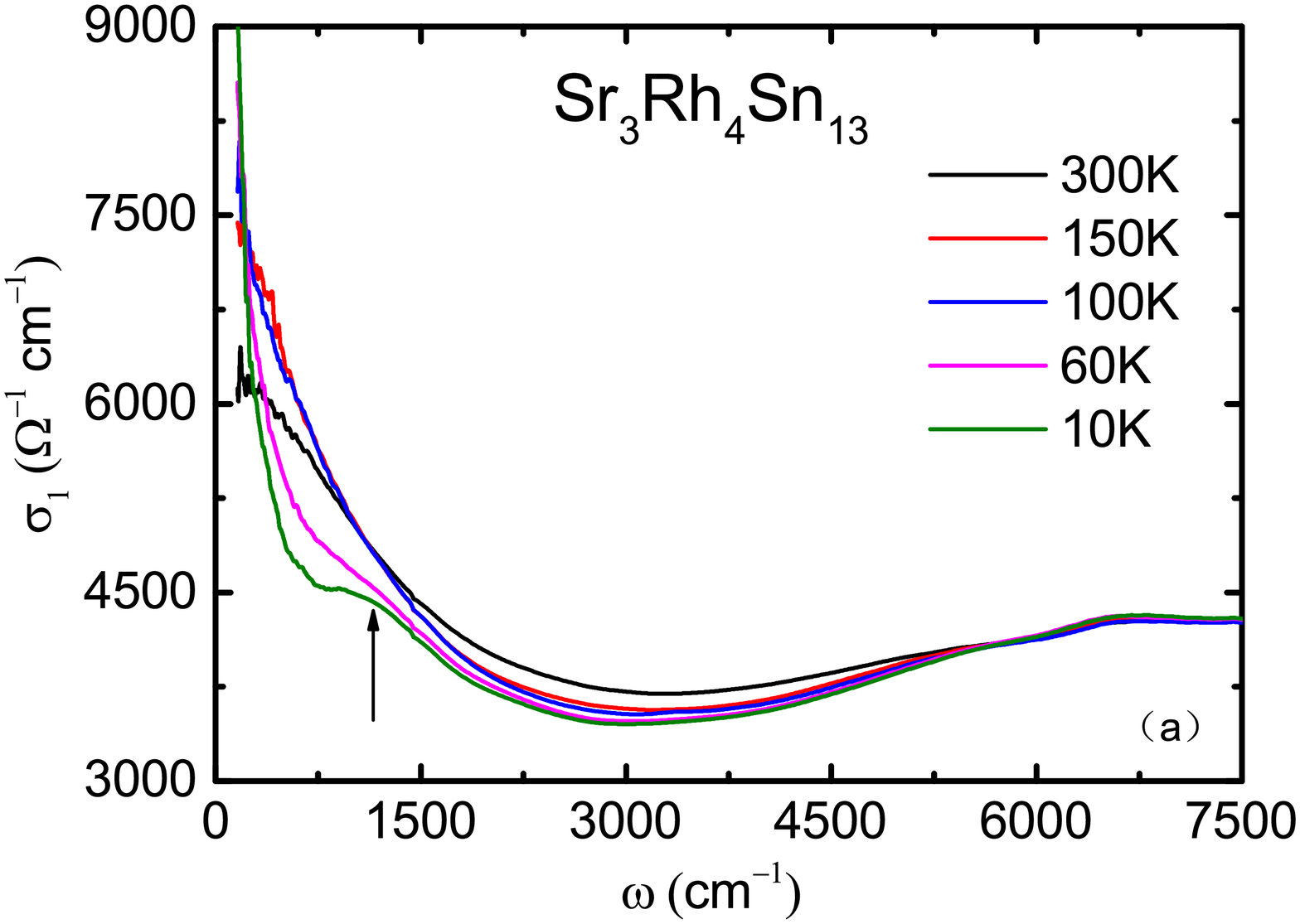}\\
  \includegraphics[width=7.5cm]{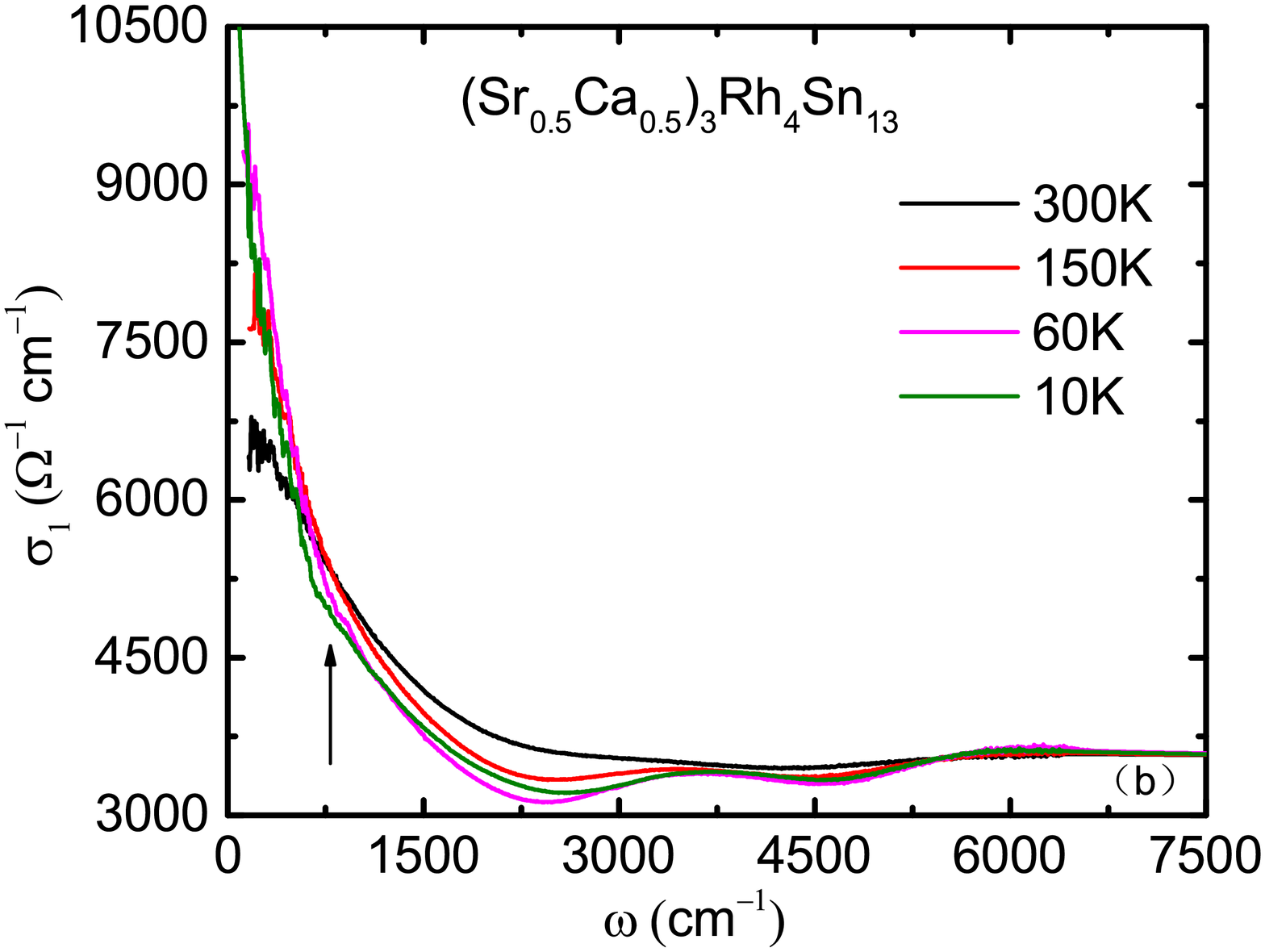}\\
  \caption{(a) The temperature dependent optical conductivity $\sigma_1(\omega)$ of Sr$_{3}$Rh$_{4}$Sn$_{13}$ (a) and (Sr$_{0.5}$Ca$_{0.5}$)$_{3}$Rh$_{4}$Sn$_{13}$ (b). The Drude component becomes narrowed at lower temperature. The arrows mark the broad peak features appeared below the  \TCDW phase transitions.}\label{Fig:conductivity}
\end{figure}

Figure \ref{Fig:conductivity} illustrates the real part of conductivity obtained by the Kramers-Kronig transformation of R($\omega$) at selected temperatures. The evolution of the electronic states is more clearly reflected in the conductivity spectra. The Drude-type conductivity is observed at all temperatures at low frequency for both Sr$_{3}$Rh$_{4}$Sn$_{13}$ and (Sr$_{0.5}$Ca$_{0.5}$)$_{3}$Rh$_{4}$Sn$_{13}$, indicating their good metallic response. For Sr$_{3}$Rh$_{4}$Sn$_{13}$, the optical conductivity between 1000 and 6000 \cm is gradually suppressed with temperature decreasing, and the Drude component becomes narrower. Upon cooling the sample across the \TCDW phase transition, a significant spectral weight suppression also develops in the Drude components and a broad peak-like feature forms near 1300 \cm which becomes more and more obvious as the temperature decreases (marked by an arrow). The existence of a residual Drude component in $\sigma_1(\omega)$ below \TCDW indicates that the Fermi surface is only partially gapped in the CDW state.

The hallmark of a symmetry-broken phase transition, such as superconductivity or density wave order, is the formation of an energy gap near the Fermi level E$_{F}$, resulting in a lowering of the total energy of the system. However, due to different coherent factors (case I for density wave and case II for superconductivity), the characteristic energy gap features of superconductivity and density wave orders are different in optical conductivity \cite{PhysRevLett.76.3838}. In an s-wave superconducting state at T = 0, the absorption smoothly rises at the gap frequency and gradually merges to the conductivity spectrum at temperatures higher than T$_c$, while for a density wave order, the opening of an energy gap leads to a nonsymmetric peak with clear edge-like feature near $2\Delta$ in the optical conductivity. The edge-like feature at $2\Delta$ could be significantly weakened for a multiple band system with presence of only a partial density wave energy gap. Nevertheless, the density wave-type energy gap value is usually estimated by using the peak position (conductivity maximum) \cite{PhysRevLett.76.3838,PhysRevLett.101.257005}. The observation of a characteristic peak-like structure above the gap yields optical evidence for a density wave type phase transition. In the (Sr$_{0.5}$Ca$_{0.5}$)$_{3}$Rh$_{4}$Sn$_{13}$ case, a similar but weaker gaplike feature is present and shifts to lower energies. So we identified the formation of a CDW induced partial energy gap in Sr$_{3}$Rh$_{4}$Sn$_{13}$ and its evolution with Ca substitution.

Since the Drude component represents the contribution from conduction electrons, the suppression of the spectral weight of Drude component observed in our optical experiment suggests the partial removal of the Fermi surfaces below the phase transition. At variance with other compounds possessing first-order structural phase transitions, e.g. IrTe$_{2}$\cite{ISI:000314359400001} or BaNi$_{2}$As$_{2}$\cite{PhysRevB.80.094506}, where the optical conductivity spectra develop sudden and dramatic changes over broad frequencies across the phase transitions owing to the reconstruction of the band structures, the present case is more similar to some CDW materials such as 2H-TaS$_{2}$ \cite{PhysRevB.76.045103}, whose spectral suppression features are rather weak and evolve continuously with temperature.

For Sr$_{3}$Rh$_{4}$Sn$_{13}$, A notable point is that the gap size at the lowest measurement temperature (10 K), which could be approximately identified as the peak position at 1300 \cm, is surprisingly large. The ratio of the energy gap relative to the transition temperature \A is about 13, And this value is almost similar for (Sr$_{0.5}$Ca$_{0.5}$)$_{3}$Rh$_{4}$Sn$_{13}$. Similar gap ratios were also observed in other low-dimensional density wave materials, e.g., the Na$_{2}$Ti$_{2}$Sb$_{2}$O\cite{PhysRevB.87.100507}, Na$_{2}$Ti$_{2}$As$_{2}$O\cite{PhysRevB.89.155120} and $Ba_{2}Ti_{2}Fe_{2}As_{4}O$\cite{PhysRevB.90.144508} compound. This ratio is much larger than the BCS (Bardeen-Cooper-Schrieffer) mean-field value of 3.52 for a density wave phase transition. This means that the transition temperature is significantly lower than the mean-field transition temperature. In low dimension compounds, the reduced phase transition temperature could be ascribed to the enhanced CDW fluctuation effect, so that the global CDW order is stabilized at lower temperature. Considering the fact that the density wave fluctuation effect should be negligible in such 3D cubic systems, the vary big values of  \A must have an unusual origin.

It is known that the conventional theory for CDW condensate is based on the weak coupling limit of electron-phonon interactions. It assumes that the coherence length $\xi_{0}$ in the transition is very long. Then the number of phonon modes which participate in the CDW phase transition is limited by this physical cutoff $k_{C}=1/\xi_{0}$. In the circumstances, the phonon frequencies are modified only over a small region of reciprocal space and the phonon entropy is unimportant as compared to the electron excitations across the gap. The weak-coupling approach predicts that the ratio of the CDW gap $\Delta$ at T =0 to transition temperature is $2\Delta/k_{B}T_{C}\approx3.52$. However, it does not fit to experimental observations in many systems, including the present observation. McMillan reformulated the theory \cite{PhysRevB.16.643}, assuming that the coherence length in the transition could be very short. This means that phonons over a substantial part of the Brillouin zone are affected in the transition so that the dominant entropy is the lattice entropy. The lattice-entropy model is in good agreement with experiment, and the short correlation length is confirmed by several experiments.

With the need for explaining the short correlation length and the consequent large observed values of $2\Delta/k_{B}T_{C}$, Varma and Simons proposed a microscopic strong coupling theory. The essential ingredients of the theory are the strong wave-vector dependence of electronically induced anharmonicity and mode-mode coupling which are shown to strongly depress the transition. This is the origin of large value of $2\Delta/k_{B}T_{C}$\cite{PhysRevLett.51.138}. In our earlier study on Sr$_{3}$Ir$_{4}$Sn$_{13}$\cite{PhysRevB.90.035115}, the strong-coupling scenario was employed to explain the very big value of  \A and the very large specific heat jump at the phase transition. This strong coupling mechanism may also apply to the present study. It deserves to remark that in the earlier study on Sr$_{3}$Ir$_{4}$Sn$_{13}$, an even larger value of  \A $\sim33$ was obtained. Here we suggest two possibilities for the observation of relative smaller values of  \A for Sr$_{3}$Rh$_{4}$Sn$_{13}$. The first one is that the Sr$_{3}$Ir$_{4}$Sn$_{13}$ has  even shorter coherence length which therefore leads to higher \A. Indeed, a comparative study on the thermodynamic properties in Sr$_{3}$Rh$_{4}$Sn$_{13}$ and Sr$_{3}$Ir$_{4}$Sn$_{13}$ by some of present authors indicated an enhancement in the electronic specific-heat jump from its mean-field value, revealing the strong-coupling nature of the observed phase transitions. Specifically, a higher power of divergence in C$_{P}$ near T$_{CDW}$ was seen for Sr$_{3}$Ir$_{4}$Sn$_{13}$ than for Sr$_{3}$Rh$_{4}$Sn$_{13}$, reflecting the fact that Sr$_{3}$Ir$_{4}$Sn$_{13}$ has a shorter coherence length \cite{PhysRevB.93.245119}. Another possibility is that in the earlier study the optical measurement was performed on polished surfaces, while in the present study the optical measurement was performed on as-grown shinny surfaces. Since mechanical polishing may induce site disorders in the surface and, to some extent, damage the sample surface, some uncertainty or extrinsic effect may be caused in the reflectance measurement. Therefore, measurement on the as-grown surface is more reliable.

\begin{figure}[htbp]
  \centering
  \includegraphics[width=7.5cm]{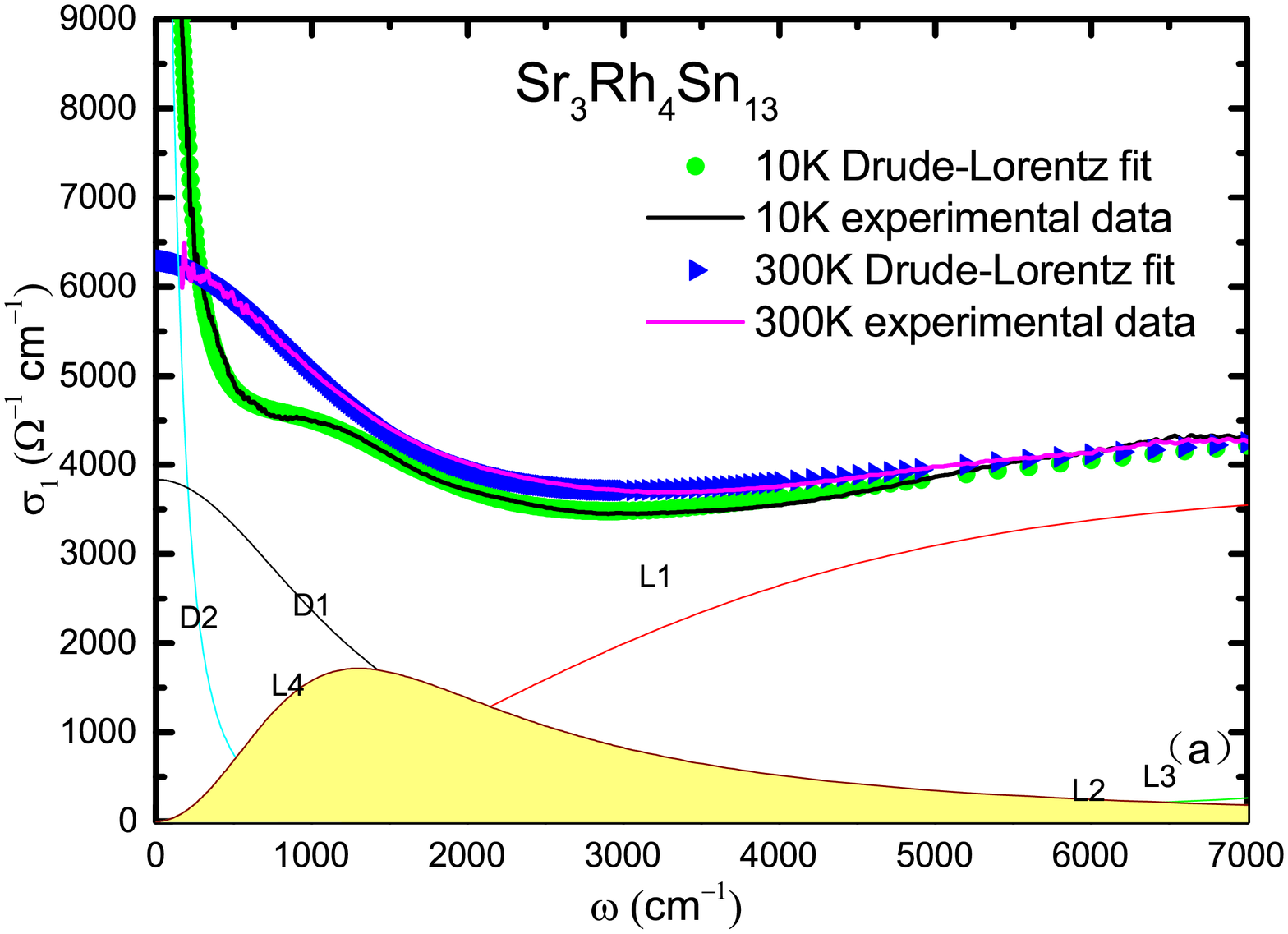}\\
  \includegraphics[width=7.5cm]{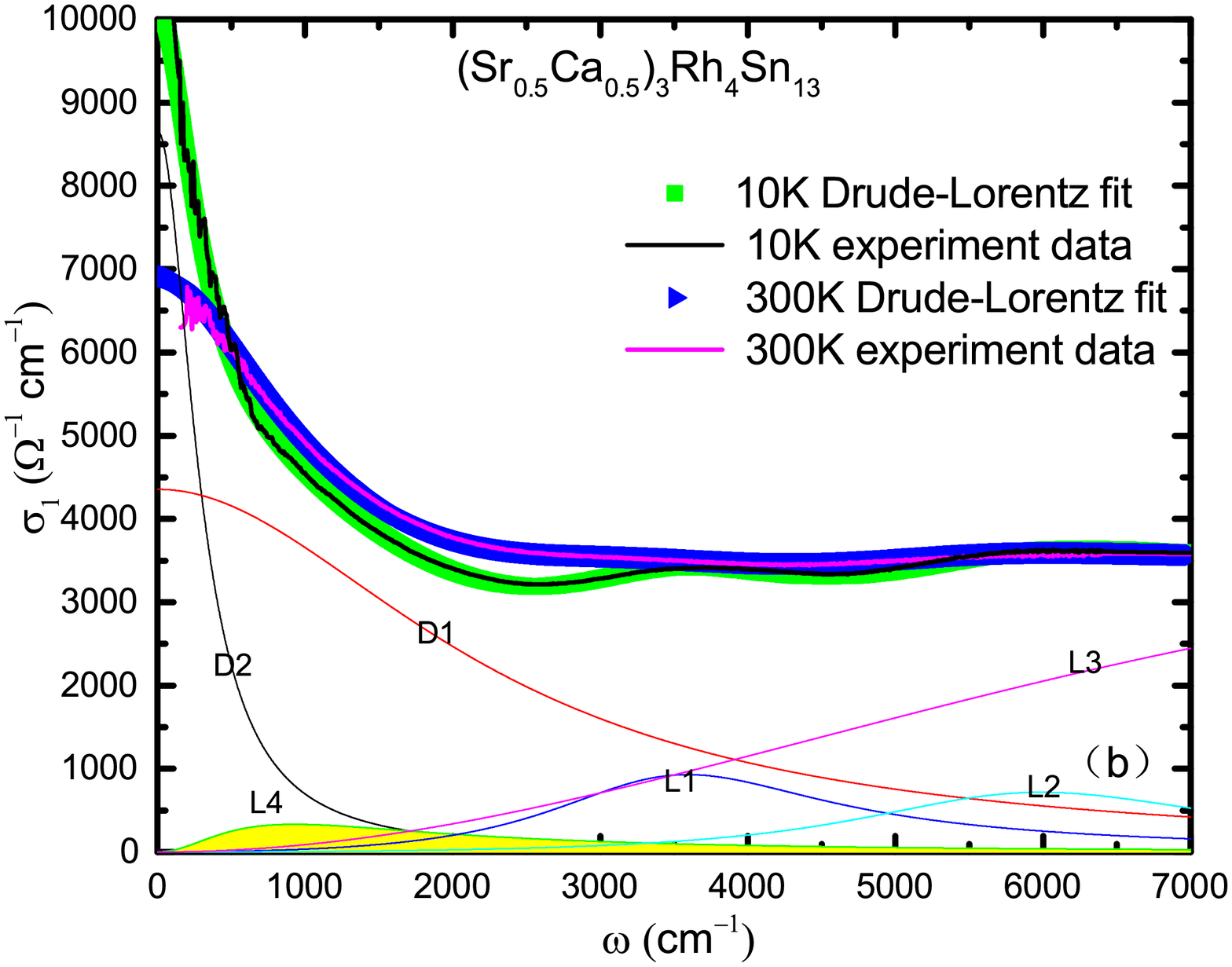}\\
  \caption{(a) The experimental optical conductivity $\sigma_1(\omega)$ along with the decomposed Drude and Lorentz components of Sr$_{3}$Rh$_{4}$Sn$_{13}$ at 10 K and 300 K (a) The experimental optical conductivity $\sigma_1(\omega)$ along with the decomposed Drude and Lorentz components of (Sr$_{0.5}$Ca$_{0.5}$)$_{3}$Rh$_{4}$Sn$_{13}$ at 10 K and 300 K.}\label{Fig:conductivity2}
\end{figure}

To quantitatively characterize the spectral change across the phase transition, particularly the evolution of the Drude part, we decompose the optical conductivity spectral into different components using a Drude-Lorentz analysis\cite{PhysRevLett.101.257005}.
\begin{equation}\label{Eq:DL}
\epsilon(\omega)= \epsilon_{\infty}-\frac{\omega_{p}^2}{\omega^2+i\omega/\tau_{D}}+ \sum_{j}{\frac{S_j^2 }{\omega_j^2-\omega^2-i\omega/\tau_j}}. 
\end{equation}
Here, $\varepsilon_{\infty}$ is the dielectric constant at high energy; the middle term is the Drude component that characterizes the electrodynamics of itinerate carriers, and the last term is the Lorentz component that describes excitations across energy gaps or interband transitions. As indicated by the band structure calculations, for Sr$_{3}$Ir$_{4}$Sn$_{13}$, several bands cross the Fermi level and the Fermi surfaces are rather complicated \cite{PhysRevB.93.235121,PhysRevLett.109.237008,PhysRevB.89.075117}. Due to the crystallographic similarity, the (Sr,Ca)$_{3}$Rh$_{4}$Sn$_{13}$ may have similar band structures. In addition, a transport study on Sr$_{3}$Rh$_{4}$Sn$_{13}$ reveals that, upon cooling, its Hall coefficient R$_{H}$ changes from negative sign to positive sign, suggests a multiband character of the Fermi surfaces \cite{PhysRevB.91.165141}. On this basis, we applied two Drude components in the analysis here. In order to reproduce the optical conductivity below 25 000 \cm at 300 K, we also add three Lorentz terms. Furthermore, an additional Lorentz component centered at 1300 \cm and 930 \cm, respectively (L4 in Fig. \ref{Fig:conductivity2}) is added at temperatures below the phase transition.

Figure \ref{Fig:conductivity2} illustrates the conductivity spectra at 10 and 300 K together with the Drude-Lorentz fitting components for Sr$_{3}$Rh$_{4}$Sn$_{13}$ and (Sr$_{0.5}$Ca$_{0.5}$)$_{3}$Rh$_{4}$Sn$_{13}$. We are mainly concerned about the evolution of the low energy conductivity, therefore the fitting parameters for the two Drude components of the two compounds are shown in the Table \ref{1} and Table \ref{2} for different temperatures. We find that the two Drude components narrow with decreasing temperature due to the metallic response. The gapping of the Fermi surfaces reduces the spectral weight of the two Drude component. Compared with the spectral weight distribution at 300 K, the spectral weight of the two Drude components decreases and the suppressed part of the two Drude components is transferred to that of the added L4 at 10 K, the other Lorentz components have nearly no change.

We use the formula  $\omega$$_{P}$=$\sqrt{\omega^{2}_{P1}+\omega^{2}_{P2}}$ to estimate the overall plasma frequency. For Sr$_{3}$Rh$_{4}$Sn$_{13}$, we obtain $\omega$$_{P}$$\approx$25950 \cm at 300 K and $\omega$$_{P}$$\approx$19770 \cm at 10 K, respectively. For (Sr$_{0.5}$Ca$_{0.5}$)$_{3}$Rh$_{4}$Sn$_{13}$, $\omega$$_{P}$$\approx$27540 \cm at 300 K and $\omega$$_{P}$$\approx$26450 \cm at 10 K. Therefore, for Sr$_{3}$Rh$_{4}$Sn$_{13}$ and (Sr$_{0.5}$Ca$_{0.5}$)$_{3}$Rh$_{4}$Sn$_{13}$, the ratio of the plasma frequency in the low temperature phase to that at 300 K is about 0.77, 0.96, respectively. It is well known that $\omega_{P}^2$ is proportional to $n/m^{*}$, where $n$ is the density of free carriers and $m^{*}$ stands for the effective mass of quasiparticles. The reduction of the plasma frequency below the phase transition could apparently be ascribed to the partial gapping of the Fermi surfaces. It also deserves remark that, although the optical measurement revealed the formation of energy gap, the measurement could not determine where the FSs are gapped. A momentum-resolved experimental probe, such as angle-resolved photoemission, should be used to determine the gapped regions and corresponding wave vectors.

Finally, we would like to remark that, for (Sr$_{0.5}$Ca$_{0.5}$)$_{3}$Rh$_{4}$Sn$_{13}$, apparent temperature induced spectral changes are also seen between 3000 and 5000 cm$^{-1}$, comparatively relatively week features are seen at higher energy scales (roughly between 5000 and 6700 cm$^{-1}$) for Sr$_{3}$Rh$_{4}$Sn$_{13}$. We would like to emphasize that those features are not directly related to the CDW order, because they are present above $T_{CDW}$ and their energy scales are much larger than the CDW energy gaps. Therefore, they should be attributed to certain interband transitions. Usually, the interband transitions involving the bands crossing the Fermi level as the initial or final states would display certain temperature dependence since the Fermi distribution function would lead to some modification on the electron distributions in those bands at different temperatures. The partial substitution of Ca for Sr should have a positive chemical pressure effect on the material. To some extent it may change the band structure of the compound, which could be the reason for the energy difference between the two samples.

\begin{table}[bhtp]
\setlength\abovecaptionskip{0.5pt}
\caption{The fitting parameters of two Drude components for Sr$_{3}$Rh$_{4}$Sn$_{13}$ at different temperatures: $\omega_{P1}$ and $\omega_{P2}$ are the plasma frequencies of the broad Drude term and the narrow Drude term, respectively, $\Gamma_{D1}$ and $\Gamma_{D2}$ are the scattering rages of the broad Drude term and the narrow Drude term, respectively. S4 is the mode strength of Lorentz 4 oscillator. The parameters of the model optical conductivity as discussed in the context is also given. The unit of these quantities is \cm. We also put the error bars for all fitting parameters in the brackets.} \label{1}
\vspace{-1em}
\begin{center}
\renewcommand\arraystretch{1.5}
\begin{tabular}{p{0.6cm} p{1.4cm} p{1.24cm} p{1.4cm} p{1.1cm} p{1.39cm} p{1.5cm}}
\hline
\hline
 T(K)&$\omega_{P1}$&$\Gamma_{D1}$&$\omega_{P2}$&$\Gamma_{D2}$&$S_{4}$&$\sqrt{\omega^{2}_{P1}+\omega^{2}_{P2}}$\\
\hline
300&19600(130)&2250(110)&17000(780)&1380(40)&0(0)&25950\\
150&19200(360)&2130(150)&17000(230)&1050(10)&0(0)&25640\\
100&17000(500)&1600(40)&13100(110)&490(10)&14400(640)&21460\\
60&17000(120)&1400(20)&10700(40)&210(10)&15500(660)&20090\\
10&17000(20)&1200(20)&10100(20)&90(5)&16300(700)&19770
\\
\hline
\hline
\end{tabular}\\
\end{center}
\end{table}

\begin{table}[bhtp]
\setlength\abovecaptionskip{0.5pt}
\caption{The fitting parameters of two Drude components for (Sr$_{0.5}$Ca$_{0.5}$)$_{3}$Rh$_{4}$Sn$_{13}$ at different temperatures.\label{2}}
\vspace{-1em}
\begin{center}
\renewcommand\arraystretch{1.5}
\begin{tabular}{p{0.6cm} p{1.4cm} p{1.21cm} p{1.4cm} p{1.1cm} p{1.23cm} p{1.5cm}}
\hline
\hline
 T(K)&$\omega_{P1}$&$\Gamma_{D1}$&$\omega_{P2}$&$\Gamma_{D2}$&$S_{4}$&$\sqrt{\omega^{2}_{P1}+\omega^{2}_{P2}}$\\
\hline
300&25100(80)&2450(8)&11340(100)&820(100)&0(0)&27540\\
150&25000(100)&2380(10)&11320(100)&560(80)&0(0)&27440\\
60&24950(80)&2300(10)&11300(100)&320(70)&0(0)&27390\\
10&24440(70)&2300(5)&10120(50)&300(5)&6510(500)&26450
\\
\hline
\hline
\end{tabular}\\
\end{center}
\end{table}

\section{conclusion}
To conclude, we perform optical spectroscopy measurements across the structural phase transition on single-crystal samples of Sr$_{3}$Rh$_{4}$Sn$_{13}$ and (Sr$_{0.5}$Ca$_{0.5}$)$_{3}$Rh$_{4}$Sn$_{13}$. Clear energy gap formations were observed for both single-crystal samples when they undergo the charge-density wave transitions. The phase transition leads to only partial removal of the itinerant carriers. The obtained 2$\Delta$/K$_{B}$T$_{CDW}$ values are roughly 13 for Sr$_{3}$Rh$_{4}$Sn$_{13}$ and (Sr$_{0.5}$Ca$_{0.5}$)$_{3}$Rh$_{4}$Sn$_{13}$. The value is considerably larger than the mean-field value based on the weak-coupling BCS theory, suggesting possibly a strong coupling mechanism. The observed spectral feature in (Sr$_{x}$Ca$_{1-x}$)$_{3}$Rh$_{4}$Sn$_{13}$ resembles those seen in many other CDW systems.

\begin{center}
\small{\textbf{ACKNOWLEDGMENTS}}
\end{center}

This work was supported by the National Science Foundation of China (No. 11327806), the National Key Research and Development Program of China (No.2016YFA0300902), and the Ministry of Science and Technology of Taiwan with the grant number of MOST-103-2112-M-006-014-MY3 (CSL).

\bibliographystyle{apsrev4-1}
  \bibliography{Sr3Rh4Sn13}

\begin{thebibliography}{38}%
\makeatletter
\providecommand \@ifxundefined [1]{%
 \@ifx{#1\undefined}
}%
\providecommand \@ifnum [1]{%
 \ifnum #1\expandafter \@firstoftwo
 \else \expandafter \@secondoftwo
 \fi
}%
\providecommand \@ifx [1]{%
 \ifx #1\expandafter \@firstoftwo
 \else \expandafter \@secondoftwo
 \fi
}%
\providecommand \natexlab [1]{#1}%
\providecommand \enquote  [1]{``#1''}%
\providecommand \bibnamefont  [1]{#1}%
\providecommand \bibfnamefont [1]{#1}%
\providecommand \citenamefont [1]{#1}%
\providecommand \href@noop [0]{\@secondoftwo}%
\providecommand \href [0]{\begingroup \@sanitize@url \@href}%
\providecommand \@href[1]{\@@startlink{#1}\@@href}%
\providecommand \@@href[1]{\endgroup#1\@@endlink}%
\providecommand \@sanitize@url [0]{\catcode `\\12\catcode `\$12\catcode
  `\&12\catcode `\#12\catcode `\^12\catcode `\_12\catcode `\%12\relax}%
\providecommand \@@startlink[1]{}%
\providecommand \@@endlink[0]{}%
\providecommand \url  [0]{\begingroup\@sanitize@url \@url }%
\providecommand \@url [1]{\endgroup\@href {#1}{\urlprefix }}%
\providecommand \urlprefix  [0]{URL }%
\providecommand \Eprint [0]{\href }%
\providecommand \doibase [0]{http://dx.doi.org/}%
\providecommand \selectlanguage [0]{\@gobble}%
\providecommand \bibinfo  [0]{\@secondoftwo}%
\providecommand \bibfield  [0]{\@secondoftwo}%
\providecommand \translation [1]{[#1]}%
\providecommand \BibitemOpen [0]{}%
\providecommand \bibitemStop [0]{}%
\providecommand \bibitemNoStop [0]{.\EOS\space}%
\providecommand \EOS [0]{\spacefactor3000\relax}%
\providecommand \BibitemShut  [1]{\csname bibitem#1\endcsname}%
\let\auto@bib@innerbib\@empty
\bibitem [{\citenamefont {Travaglini}\ and\ \citenamefont
  {Wachter}(1984)}]{PhysRevB.30.1971}%
  \BibitemOpen
  \bibfield  {author} {\bibinfo {author} {\bibfnamefont {G.}~\bibnamefont
  {Travaglini}}\ and\ \bibinfo {author} {\bibfnamefont {P.}~\bibnamefont
  {Wachter}},\ }\href {\doibase 10.1103/PhysRevB.30.1971} {\bibfield  {journal}
  {\bibinfo  {journal} {Phys. Rev. B}\ }\textbf {\bibinfo {volume} {30}},\
  \bibinfo {pages} {1971} (\bibinfo {year} {1984})}\BibitemShut {NoStop}%
\bibitem [{\citenamefont {WU}\ and\ \citenamefont
  {LIEBER}(1989)}]{ISI:A1989T890400023}%
  \BibitemOpen
  \bibfield  {author} {\bibinfo {author} {\bibfnamefont {X.~L.}\ \bibnamefont
  {WU}}\ and\ \bibinfo {author} {\bibfnamefont {C.~M.}\ \bibnamefont
  {LIEBER}},\ }\href {\doibase 10.1126/science.243.4899.1703} {\bibfield
  {journal} {\bibinfo  {journal} {SCIENCE}\ }\textbf {\bibinfo {volume}
  {243}},\ \bibinfo {pages} {1703} (\bibinfo {year} {1989})}\BibitemShut
  {NoStop}%
\bibitem [{\citenamefont {Wang}\ \emph {et~al.}(2015)\citenamefont {Wang},
  \citenamefont {Wang}, \citenamefont {Chen}, \citenamefont {Kuo},\ and\
  \citenamefont {Lue}}]{1367-2630-17-3-033005}%
  \BibitemOpen
  \bibfield  {author} {\bibinfo {author} {\bibfnamefont {L.~M.}\ \bibnamefont
  {Wang}}, \bibinfo {author} {\bibfnamefont {C.-Y.}\ \bibnamefont {Wang}},
  \bibinfo {author} {\bibfnamefont {G.-M.}\ \bibnamefont {Chen}}, \bibinfo
  {author} {\bibfnamefont {C.~N.}\ \bibnamefont {Kuo}}, \ and\ \bibinfo
  {author} {\bibfnamefont {C.~S.}\ \bibnamefont {Lue}},\ }\href
  {http://stacks.iop.org/1367-2630/17/i=3/a=033005} {\bibfield  {journal}
  {\bibinfo  {journal} {New Journal of Physics}\ }\textbf {\bibinfo {volume}
  {17}},\ \bibinfo {pages} {033005} (\bibinfo {year} {2015})}\BibitemShut
  {NoStop}%
\bibitem [{\citenamefont {Mazzone}\ \emph {et~al.}(2015)\citenamefont
  {Mazzone}, \citenamefont {Gerber}, \citenamefont {Gavilano}, \citenamefont
  {Sibille}, \citenamefont {Medarde}, \citenamefont {Delley}, \citenamefont
  {Ramakrishnan}, \citenamefont {Neugebauer}, \citenamefont {Regnault},
  \citenamefont {Chernyshov}, \citenamefont {Piovano}, \citenamefont
  {Fernandez-Diaz}, \citenamefont {Keller}, \citenamefont {Cervellino},
  \citenamefont {Pomjakushina}, \citenamefont {Conder},\ and\ \citenamefont
  {Kenzelmann}}]{PhysRevB.92.024101}%
  \BibitemOpen
  \bibfield  {author} {\bibinfo {author} {\bibfnamefont {D.~G.}\ \bibnamefont
  {Mazzone}}, \bibinfo {author} {\bibfnamefont {S.}~\bibnamefont {Gerber}},
  \bibinfo {author} {\bibfnamefont {J.~L.}\ \bibnamefont {Gavilano}}, \bibinfo
  {author} {\bibfnamefont {R.}~\bibnamefont {Sibille}}, \bibinfo {author}
  {\bibfnamefont {M.}~\bibnamefont {Medarde}}, \bibinfo {author} {\bibfnamefont
  {B.}~\bibnamefont {Delley}}, \bibinfo {author} {\bibfnamefont
  {M.}~\bibnamefont {Ramakrishnan}}, \bibinfo {author} {\bibfnamefont
  {M.}~\bibnamefont {Neugebauer}}, \bibinfo {author} {\bibfnamefont {L.~P.}\
  \bibnamefont {Regnault}}, \bibinfo {author} {\bibfnamefont {D.}~\bibnamefont
  {Chernyshov}}, \bibinfo {author} {\bibfnamefont {A.}~\bibnamefont {Piovano}},
  \bibinfo {author} {\bibfnamefont {T.~M.}\ \bibnamefont {Fernandez-Diaz}},
  \bibinfo {author} {\bibfnamefont {L.}~\bibnamefont {Keller}}, \bibinfo
  {author} {\bibfnamefont {A.}~\bibnamefont {Cervellino}}, \bibinfo {author}
  {\bibfnamefont {E.}~\bibnamefont {Pomjakushina}}, \bibinfo {author}
  {\bibfnamefont {K.}~\bibnamefont {Conder}}, \ and\ \bibinfo {author}
  {\bibfnamefont {M.}~\bibnamefont {Kenzelmann}},\ }\href {\doibase
  10.1103/PhysRevB.92.024101} {\bibfield  {journal} {\bibinfo  {journal} {Phys.
  Rev. B}\ }\textbf {\bibinfo {volume} {92}},\ \bibinfo {pages} {024101}
  (\bibinfo {year} {2015})}\BibitemShut {NoStop}%
\bibitem [{\citenamefont {Biswas}\ \emph {et~al.}(2015)\citenamefont {Biswas},
  \citenamefont {Guguchia}, \citenamefont {Khasanov}, \citenamefont {Chinotti},
  \citenamefont {Li}, \citenamefont {Wang}, \citenamefont {Petrovic},\ and\
  \citenamefont {Morenzoni}}]{PhysRevB.92.195122}%
  \BibitemOpen
  \bibfield  {author} {\bibinfo {author} {\bibfnamefont {P.~K.}\ \bibnamefont
  {Biswas}}, \bibinfo {author} {\bibfnamefont {Z.}~\bibnamefont {Guguchia}},
  \bibinfo {author} {\bibfnamefont {R.}~\bibnamefont {Khasanov}}, \bibinfo
  {author} {\bibfnamefont {M.}~\bibnamefont {Chinotti}}, \bibinfo {author}
  {\bibfnamefont {L.}~\bibnamefont {Li}}, \bibinfo {author} {\bibfnamefont
  {K.}~\bibnamefont {Wang}}, \bibinfo {author} {\bibfnamefont {C.}~\bibnamefont
  {Petrovic}}, \ and\ \bibinfo {author} {\bibfnamefont {E.}~\bibnamefont
  {Morenzoni}},\ }\href {\doibase 10.1103/PhysRevB.92.195122} {\bibfield
  {journal} {\bibinfo  {journal} {Phys. Rev. B}\ }\textbf {\bibinfo {volume}
  {92}},\ \bibinfo {pages} {195122} (\bibinfo {year} {2015})}\BibitemShut
  {NoStop}%
\bibitem [{\citenamefont {Hou}\ \emph {et~al.}(2016)\citenamefont {Hou},
  \citenamefont {Wong}, \citenamefont {Lortz}, \citenamefont {Sibille},\ and\
  \citenamefont {Kenzelmann}}]{PhysRevB.93.134505}%
  \BibitemOpen
  \bibfield  {author} {\bibinfo {author} {\bibfnamefont {J.}~\bibnamefont
  {Hou}}, \bibinfo {author} {\bibfnamefont {C.~H.}\ \bibnamefont {Wong}},
  \bibinfo {author} {\bibfnamefont {R.}~\bibnamefont {Lortz}}, \bibinfo
  {author} {\bibfnamefont {R.}~\bibnamefont {Sibille}}, \ and\ \bibinfo
  {author} {\bibfnamefont {M.}~\bibnamefont {Kenzelmann}},\ }\href {\doibase
  10.1103/PhysRevB.93.134505} {\bibfield  {journal} {\bibinfo  {journal} {Phys.
  Rev. B}\ }\textbf {\bibinfo {volume} {93}},\ \bibinfo {pages} {134505}
  (\bibinfo {year} {2016})}\BibitemShut {NoStop}%
\bibitem [{\citenamefont {Lue}\ \emph {et~al.}(2016)\citenamefont {Lue},
  \citenamefont {Kuo}, \citenamefont {Tseng}, \citenamefont {Wu}, \citenamefont
  {Liang}, \citenamefont {Du},\ and\ \citenamefont {Kuo}}]{PhysRevB.93.245119}%
  \BibitemOpen
  \bibfield  {author} {\bibinfo {author} {\bibfnamefont {C.~S.}\ \bibnamefont
  {Lue}}, \bibinfo {author} {\bibfnamefont {C.~N.}\ \bibnamefont {Kuo}},
  \bibinfo {author} {\bibfnamefont {C.~W.}\ \bibnamefont {Tseng}}, \bibinfo
  {author} {\bibfnamefont {K.~K.}\ \bibnamefont {Wu}}, \bibinfo {author}
  {\bibfnamefont {Y.-H.}\ \bibnamefont {Liang}}, \bibinfo {author}
  {\bibfnamefont {C.-H.}\ \bibnamefont {Du}}, \ and\ \bibinfo {author}
  {\bibfnamefont {Y.~K.}\ \bibnamefont {Kuo}},\ }\href {\doibase
  10.1103/PhysRevB.93.245119} {\bibfield  {journal} {\bibinfo  {journal} {Phys.
  Rev. B}\ }\textbf {\bibinfo {volume} {93}},\ \bibinfo {pages} {245119}
  (\bibinfo {year} {2016})}\BibitemShut {NoStop}%
\bibitem [{\citenamefont {\ifmmode~\acute{S}\else \'{S}\fi{}lebarski}\ \emph
  {et~al.}(2016)\citenamefont {\ifmmode~\acute{S}\else \'{S}\fi{}lebarski},
  \citenamefont {Goraus}, \citenamefont {Ma\ifmmode~\acute{s}\else
  \'{s}\fi{}ka}, \citenamefont {Witas}, \citenamefont {Fija\l{}kowski},
  \citenamefont {Wolowiec}, \citenamefont {Fang},\ and\ \citenamefont
  {Maple}}]{PhysRevB.93.245126}%
  \BibitemOpen
  \bibfield  {author} {\bibinfo {author} {\bibfnamefont {A.}~\bibnamefont
  {\ifmmode~\acute{S}\else \'{S}\fi{}lebarski}}, \bibinfo {author}
  {\bibfnamefont {J.}~\bibnamefont {Goraus}}, \bibinfo {author} {\bibfnamefont
  {M.~M.}\ \bibnamefont {Ma\ifmmode~\acute{s}\else \'{s}\fi{}ka}}, \bibinfo
  {author} {\bibfnamefont {P.}~\bibnamefont {Witas}}, \bibinfo {author}
  {\bibfnamefont {M.}~\bibnamefont {Fija\l{}kowski}}, \bibinfo {author}
  {\bibfnamefont {C.~T.}\ \bibnamefont {Wolowiec}}, \bibinfo {author}
  {\bibfnamefont {Y.}~\bibnamefont {Fang}}, \ and\ \bibinfo {author}
  {\bibfnamefont {M.~B.}\ \bibnamefont {Maple}},\ }\href {\doibase
  10.1103/PhysRevB.93.245126} {\bibfield  {journal} {\bibinfo  {journal}
  {Phys.rev.B}\ }\textbf {\bibinfo {volume} {93}},\ \bibinfo {pages} {245126}
  (\bibinfo {year} {2016})}\BibitemShut {NoStop}%
\bibitem [{\citenamefont {C.W.Luo}\ \emph {et~al.}(2016)\citenamefont
  {C.W.Luo}, \citenamefont {P.C.Cheng}, \citenamefont {C.M.Tu}, \citenamefont
  {C.N.Kuo}, \citenamefont {C.M.Wang},\ and\ \citenamefont
  {C.S.Lue}}]{1367-2630-18-7-073045}%
  \BibitemOpen
  \bibfield  {author} {\bibinfo {author} {\bibnamefont {C.W.Luo}}, \bibinfo
  {author} {\bibnamefont {P.C.Cheng}}, \bibinfo {author} {\bibnamefont
  {C.M.Tu}}, \bibinfo {author} {\bibnamefont {C.N.Kuo}}, \bibinfo {author}
  {\bibnamefont {C.M.Wang}}, \ and\ \bibinfo {author} {\bibnamefont
  {C.S.Lue}},\ }\href {http://stacks.iop.org/1367-2630/18/i=7/a=073045}
  {\bibfield  {journal} {\bibinfo  {journal} {New Journal of Physics}\ }\textbf
  {\bibinfo {volume} {18}},\ \bibinfo {pages} {073045} (\bibinfo {year}
  {2016})}\BibitemShut {NoStop}%
\bibitem [{\citenamefont {Yang}\ \emph {et~al.}(2010)\citenamefont {Yang},
  \citenamefont {Chen}, \citenamefont {Michioka},\ and\ \citenamefont
  {Yoshimura}}]{doi:10.1143/JPSJ.79.113705}%
  \BibitemOpen
  \bibfield  {author} {\bibinfo {author} {\bibfnamefont {J.}~\bibnamefont
  {Yang}}, \bibinfo {author} {\bibfnamefont {B.}~\bibnamefont {Chen}}, \bibinfo
  {author} {\bibfnamefont {C.}~\bibnamefont {Michioka}}, \ and\ \bibinfo
  {author} {\bibfnamefont {K.}~\bibnamefont {Yoshimura}},\ }\href {\doibase
  10.1143/JPSJ.79.113705} {\bibfield  {journal} {\bibinfo  {journal} {Journal
  of the Physical Society of Japan}\ }\textbf {\bibinfo {volume} {79}},\
  \bibinfo {pages} {113705} (\bibinfo {year} {2010})}\BibitemShut {NoStop}%
\bibitem [{\citenamefont {Kase}\ \emph {et~al.}(2011)\citenamefont {Kase},
  \citenamefont {Hayamizu},\ and\ \citenamefont
  {Akimitsu}}]{PhysRevB.83.184509}%
  \BibitemOpen
  \bibfield  {author} {\bibinfo {author} {\bibfnamefont {N.}~\bibnamefont
  {Kase}}, \bibinfo {author} {\bibfnamefont {H.}~\bibnamefont {Hayamizu}}, \
  and\ \bibinfo {author} {\bibfnamefont {J.}~\bibnamefont {Akimitsu}},\ }\href
  {\doibase 10.1103/PhysRevB.83.184509} {\bibfield  {journal} {\bibinfo
  {journal} {Phys. Rev. B}\ }\textbf {\bibinfo {volume} {83}},\ \bibinfo
  {pages} {184509} (\bibinfo {year} {2011})}\BibitemShut {NoStop}%
\bibitem [{\citenamefont {Hayamizu}\ \emph {et~al.}(2011)\citenamefont
  {Hayamizu}, \citenamefont {Kase},\ and\ \citenamefont
  {Akimitsu}}]{doi:10.1143/JPSJS.80SA.SA114}%
  \BibitemOpen
  \bibfield  {author} {\bibinfo {author} {\bibfnamefont {H.}~\bibnamefont
  {Hayamizu}}, \bibinfo {author} {\bibfnamefont {N.}~\bibnamefont {Kase}}, \
  and\ \bibinfo {author} {\bibfnamefont {J.}~\bibnamefont {Akimitsu}},\ }\href
  {\doibase 10.1143/JPSJS.80SA.SA114} {\bibfield  {journal} {\bibinfo
  {journal} {Journal of the Physical Society of Japan}\ }\textbf {\bibinfo
  {volume} {80}},\ \bibinfo {pages} {SA114} (\bibinfo {year}
  {2011})}\BibitemShut {NoStop}%
\bibitem [{\citenamefont {Wang}\ and\ \citenamefont
  {Petrovic}(2012)}]{PhysRevB.86.024522}%
  \BibitemOpen
  \bibfield  {author} {\bibinfo {author} {\bibfnamefont {K.}~\bibnamefont
  {Wang}}\ and\ \bibinfo {author} {\bibfnamefont {C.}~\bibnamefont
  {Petrovic}},\ }\href {\doibase 10.1103/PhysRevB.86.024522} {\bibfield
  {journal} {\bibinfo  {journal} {Phys. Rev. B}\ }\textbf {\bibinfo {volume}
  {86}},\ \bibinfo {pages} {024522} (\bibinfo {year} {2012})}\BibitemShut
  {NoStop}%
\bibitem [{\citenamefont {Zhou}\ \emph {et~al.}(2012)\citenamefont {Zhou},
  \citenamefont {Zhang}, \citenamefont {Hong}, \citenamefont {Pan},
  \citenamefont {Qiu}, \citenamefont {Dong}, \citenamefont {Li},\ and\
  \citenamefont {Li}}]{PhysRevB.86.064504}%
  \BibitemOpen
  \bibfield  {author} {\bibinfo {author} {\bibfnamefont {S.~Y.}\ \bibnamefont
  {Zhou}}, \bibinfo {author} {\bibfnamefont {H.}~\bibnamefont {Zhang}},
  \bibinfo {author} {\bibfnamefont {X.~C.}\ \bibnamefont {Hong}}, \bibinfo
  {author} {\bibfnamefont {B.~Y.}\ \bibnamefont {Pan}}, \bibinfo {author}
  {\bibfnamefont {X.}~\bibnamefont {Qiu}}, \bibinfo {author} {\bibfnamefont
  {W.~N.}\ \bibnamefont {Dong}}, \bibinfo {author} {\bibfnamefont {X.~L.}\
  \bibnamefont {Li}}, \ and\ \bibinfo {author} {\bibfnamefont {S.~Y.}\
  \bibnamefont {Li}},\ }\href {\doibase 10.1103/PhysRevB.86.064504} {\bibfield
  {journal} {\bibinfo  {journal} {Phys. Rev. B}\ }\textbf {\bibinfo {volume}
  {86}},\ \bibinfo {pages} {064504} (\bibinfo {year} {2012})}\BibitemShut
  {NoStop}%
\bibitem [{\citenamefont {Gerber}\ \emph {et~al.}(2013)\citenamefont {Gerber},
  \citenamefont {Gavilano}, \citenamefont {Medarde}, \citenamefont
  {Pomjakushin}, \citenamefont {Baines}, \citenamefont {Pomjakushina},
  \citenamefont {Conder},\ and\ \citenamefont
  {Kenzelmann}}]{PhysRevB.88.104505}%
  \BibitemOpen
  \bibfield  {author} {\bibinfo {author} {\bibfnamefont {S.}~\bibnamefont
  {Gerber}}, \bibinfo {author} {\bibfnamefont {J.~L.}\ \bibnamefont
  {Gavilano}}, \bibinfo {author} {\bibfnamefont {M.}~\bibnamefont {Medarde}},
  \bibinfo {author} {\bibfnamefont {V.}~\bibnamefont {Pomjakushin}}, \bibinfo
  {author} {\bibfnamefont {C.}~\bibnamefont {Baines}}, \bibinfo {author}
  {\bibfnamefont {E.}~\bibnamefont {Pomjakushina}}, \bibinfo {author}
  {\bibfnamefont {K.}~\bibnamefont {Conder}}, \ and\ \bibinfo {author}
  {\bibfnamefont {M.}~\bibnamefont {Kenzelmann}},\ }\href {\doibase
  10.1103/PhysRevB.88.104505} {\bibfield  {journal} {\bibinfo  {journal} {Phys.
  Rev. B}\ }\textbf {\bibinfo {volume} {88}},\ \bibinfo {pages} {104505}
  (\bibinfo {year} {2013})}\BibitemShut {NoStop}%
\bibitem [{\citenamefont {Liu}\ \emph {et~al.}(2013)\citenamefont {Liu},
  \citenamefont {Kuo}, \citenamefont {Lue}, \citenamefont {Syu},\ and\
  \citenamefont {Kuo}}]{PhysRevB.88.115113}%
  \BibitemOpen
  \bibfield  {author} {\bibinfo {author} {\bibfnamefont {H.~F.}\ \bibnamefont
  {Liu}}, \bibinfo {author} {\bibfnamefont {C.~N.}\ \bibnamefont {Kuo}},
  \bibinfo {author} {\bibfnamefont {C.~S.}\ \bibnamefont {Lue}}, \bibinfo
  {author} {\bibfnamefont {K.-Z.}\ \bibnamefont {Syu}}, \ and\ \bibinfo
  {author} {\bibfnamefont {Y.~K.}\ \bibnamefont {Kuo}},\ }\href {\doibase
  10.1103/PhysRevB.88.115113} {\bibfield  {journal} {\bibinfo  {journal} {Phys.
  Rev. B}\ }\textbf {\bibinfo {volume} {88}},\ \bibinfo {pages} {115113}
  (\bibinfo {year} {2013})}\BibitemShut {NoStop}%
\bibitem [{\citenamefont {\ifmmode~\acute{S}\else \'{S}\fi{}lebarski}\ \emph
  {et~al.}(2014)\citenamefont {\ifmmode~\acute{S}\else \'{S}\fi{}lebarski},
  \citenamefont {Fija\l{}kowski}, \citenamefont {Ma\ifmmode~\acute{s}\else
  \'{s}\fi{}ka}, \citenamefont {Mierzejewski}, \citenamefont {White},\ and\
  \citenamefont {Maple}}]{PhysRevB.89.125111}%
  \BibitemOpen
  \bibfield  {author} {\bibinfo {author} {\bibfnamefont {A.}~\bibnamefont
  {\ifmmode~\acute{S}\else \'{S}\fi{}lebarski}}, \bibinfo {author}
  {\bibfnamefont {M.}~\bibnamefont {Fija\l{}kowski}}, \bibinfo {author}
  {\bibfnamefont {M.~M.}\ \bibnamefont {Ma\ifmmode~\acute{s}\else
  \'{s}\fi{}ka}}, \bibinfo {author} {\bibfnamefont {M.}~\bibnamefont
  {Mierzejewski}}, \bibinfo {author} {\bibfnamefont {B.~D.}\ \bibnamefont
  {White}}, \ and\ \bibinfo {author} {\bibfnamefont {M.~B.}\ \bibnamefont
  {Maple}},\ }\href {\doibase 10.1103/PhysRevB.89.125111} {\bibfield  {journal}
  {\bibinfo  {journal} {Phys. Rev. B}\ }\textbf {\bibinfo {volume} {89}},\
  \bibinfo {pages} {125111} (\bibinfo {year} {2014})}\BibitemShut {NoStop}%
\bibitem [{\citenamefont {Yu}\ \emph {et~al.}(2015)\citenamefont {Yu},
  \citenamefont {Cheung}, \citenamefont {Saines}, \citenamefont {Imai},
  \citenamefont {Matsumoto}, \citenamefont {Michioka}, \citenamefont
  {Yoshimura},\ and\ \citenamefont {Goh}}]{PhysRevLett.115.207003}%
  \BibitemOpen
  \bibfield  {author} {\bibinfo {author} {\bibfnamefont {W.~C.}\ \bibnamefont
  {Yu}}, \bibinfo {author} {\bibfnamefont {Y.~W.}\ \bibnamefont {Cheung}},
  \bibinfo {author} {\bibfnamefont {P.~J.}\ \bibnamefont {Saines}}, \bibinfo
  {author} {\bibfnamefont {M.}~\bibnamefont {Imai}}, \bibinfo {author}
  {\bibfnamefont {T.}~\bibnamefont {Matsumoto}}, \bibinfo {author}
  {\bibfnamefont {C.}~\bibnamefont {Michioka}}, \bibinfo {author}
  {\bibfnamefont {K.}~\bibnamefont {Yoshimura}}, \ and\ \bibinfo {author}
  {\bibfnamefont {S.~K.}\ \bibnamefont {Goh}},\ }\href {\doibase
  10.1103/PhysRevLett.115.207003} {\bibfield  {journal} {\bibinfo  {journal}
  {Phys. Rev. Lett.}\ }\textbf {\bibinfo {volume} {115}},\ \bibinfo {pages}
  {207003} (\bibinfo {year} {2015})}\BibitemShut {NoStop}%
\bibitem [{\citenamefont {Tompsett}(2014)}]{PhysRevB.89.075117}%
  \BibitemOpen
  \bibfield  {author} {\bibinfo {author} {\bibfnamefont {D.~A.}\ \bibnamefont
  {Tompsett}},\ }\href {\doibase 10.1103/PhysRevB.89.075117} {\bibfield
  {journal} {\bibinfo  {journal} {Phys. Rev. B}\ }\textbf {\bibinfo {volume}
  {89}},\ \bibinfo {pages} {075117} (\bibinfo {year} {2014})}\BibitemShut
  {NoStop}%
\bibitem [{\citenamefont {Biswas}\ \emph {et~al.}(2014)\citenamefont {Biswas},
  \citenamefont {Amato}, \citenamefont {Khasanov}, \citenamefont {Luetkens},
  \citenamefont {Wang}, \citenamefont {Petrovic}, \citenamefont {Cook},
  \citenamefont {Lees},\ and\ \citenamefont {Morenzoni}}]{PhysRevB.90.144505}%
  \BibitemOpen
  \bibfield  {author} {\bibinfo {author} {\bibfnamefont {P.~K.}\ \bibnamefont
  {Biswas}}, \bibinfo {author} {\bibfnamefont {A.}~\bibnamefont {Amato}},
  \bibinfo {author} {\bibfnamefont {R.}~\bibnamefont {Khasanov}}, \bibinfo
  {author} {\bibfnamefont {H.}~\bibnamefont {Luetkens}}, \bibinfo {author}
  {\bibfnamefont {K.}~\bibnamefont {Wang}}, \bibinfo {author} {\bibfnamefont
  {C.}~\bibnamefont {Petrovic}}, \bibinfo {author} {\bibfnamefont {R.~M.}\
  \bibnamefont {Cook}}, \bibinfo {author} {\bibfnamefont {M.~R.}\ \bibnamefont
  {Lees}}, \ and\ \bibinfo {author} {\bibfnamefont {E.}~\bibnamefont
  {Morenzoni}},\ }\href {\doibase 10.1103/PhysRevB.90.144505} {\bibfield
  {journal} {\bibinfo  {journal} {Phys. Rev. B}\ }\textbf {\bibinfo {volume}
  {90}},\ \bibinfo {pages} {144505} (\bibinfo {year} {2014})}\BibitemShut
  {NoStop}%
\bibitem [{\citenamefont {Fang}\ \emph {et~al.}(2014)\citenamefont {Fang},
  \citenamefont {Wang}, \citenamefont {Zheng},\ and\ \citenamefont
  {Wang}}]{PhysRevB.90.035115}%
  \BibitemOpen
  \bibfield  {author} {\bibinfo {author} {\bibfnamefont {A.~F.}\ \bibnamefont
  {Fang}}, \bibinfo {author} {\bibfnamefont {X.~B.}\ \bibnamefont {Wang}},
  \bibinfo {author} {\bibfnamefont {P.}~\bibnamefont {Zheng}}, \ and\ \bibinfo
  {author} {\bibfnamefont {N.~L.}\ \bibnamefont {Wang}},\ }\href {\doibase
  10.1103/PhysRevB.90.035115} {\bibfield  {journal} {\bibinfo  {journal} {Phys.
  Rev. B}\ }\textbf {\bibinfo {volume} {90}},\ \bibinfo {pages} {035115}
  (\bibinfo {year} {2014})}\BibitemShut {NoStop}%
\bibitem [{\citenamefont {Kuo}\ \emph {et~al.}(2014)\citenamefont {Kuo},
  \citenamefont {Liu}, \citenamefont {Lue}, \citenamefont {Wang}, \citenamefont
  {Chen},\ and\ \citenamefont {Kuo}}]{PhysRevB.89.094520}%
  \BibitemOpen
  \bibfield  {author} {\bibinfo {author} {\bibfnamefont {C.~N.}\ \bibnamefont
  {Kuo}}, \bibinfo {author} {\bibfnamefont {H.~F.}\ \bibnamefont {Liu}},
  \bibinfo {author} {\bibfnamefont {C.~S.}\ \bibnamefont {Lue}}, \bibinfo
  {author} {\bibfnamefont {L.~M.}\ \bibnamefont {Wang}}, \bibinfo {author}
  {\bibfnamefont {C.~C.}\ \bibnamefont {Chen}}, \ and\ \bibinfo {author}
  {\bibfnamefont {Y.~K.}\ \bibnamefont {Kuo}},\ }\href {\doibase
  10.1103/PhysRevB.89.094520} {\bibfield  {journal} {\bibinfo  {journal} {Phys.
  Rev. B}\ }\textbf {\bibinfo {volume} {89}},\ \bibinfo {pages} {094520}
  (\bibinfo {year} {2014})}\BibitemShut {NoStop}%
\bibitem [{\citenamefont {Goh}\ \emph {et~al.}(2015)\citenamefont {Goh},
  \citenamefont {Tompsett}, \citenamefont {Saines}, \citenamefont {Chang},
  \citenamefont {Matsumoto}, \citenamefont {Imai}, \citenamefont {Yoshimura},\
  and\ \citenamefont {Grosche}}]{PhysRevLett.114.097002}%
  \BibitemOpen
  \bibfield  {author} {\bibinfo {author} {\bibfnamefont {S.~K.}\ \bibnamefont
  {Goh}}, \bibinfo {author} {\bibfnamefont {D.~A.}\ \bibnamefont {Tompsett}},
  \bibinfo {author} {\bibfnamefont {P.~J.}\ \bibnamefont {Saines}}, \bibinfo
  {author} {\bibfnamefont {H.~C.}\ \bibnamefont {Chang}}, \bibinfo {author}
  {\bibfnamefont {T.}~\bibnamefont {Matsumoto}}, \bibinfo {author}
  {\bibfnamefont {M.}~\bibnamefont {Imai}}, \bibinfo {author} {\bibfnamefont
  {K.}~\bibnamefont {Yoshimura}}, \ and\ \bibinfo {author} {\bibfnamefont
  {F.~M.}\ \bibnamefont {Grosche}},\ }\href {\doibase
  10.1103/PhysRevLett.114.097002} {\bibfield  {journal} {\bibinfo  {journal}
  {Phys. Rev. Lett.}\ }\textbf {\bibinfo {volume} {114}},\ \bibinfo {pages}
  {097002} (\bibinfo {year} {2015})}\BibitemShut {NoStop}%
\bibitem [{\citenamefont {Klintberg}\ \emph {et~al.}(2012)\citenamefont
  {Klintberg}, \citenamefont {Goh}, \citenamefont {Alireza}, \citenamefont
  {Saines}, \citenamefont {Tompsett}, \citenamefont {Logg}, \citenamefont
  {Yang}, \citenamefont {Chen}, \citenamefont {Yoshimura},\ and\ \citenamefont
  {Grosche}}]{PhysRevLett.109.237008}%
  \BibitemOpen
  \bibfield  {author} {\bibinfo {author} {\bibfnamefont {L.~E.}\ \bibnamefont
  {Klintberg}}, \bibinfo {author} {\bibfnamefont {S.~K.}\ \bibnamefont {Goh}},
  \bibinfo {author} {\bibfnamefont {P.~L.}\ \bibnamefont {Alireza}}, \bibinfo
  {author} {\bibfnamefont {P.~J.}\ \bibnamefont {Saines}}, \bibinfo {author}
  {\bibfnamefont {D.~A.}\ \bibnamefont {Tompsett}}, \bibinfo {author}
  {\bibfnamefont {P.~W.}\ \bibnamefont {Logg}}, \bibinfo {author}
  {\bibfnamefont {J.}~\bibnamefont {Yang}}, \bibinfo {author} {\bibfnamefont
  {B.}~\bibnamefont {Chen}}, \bibinfo {author} {\bibfnamefont {K.}~\bibnamefont
  {Yoshimura}}, \ and\ \bibinfo {author} {\bibfnamefont {F.~M.}\ \bibnamefont
  {Grosche}},\ }\href {\doibase 10.1103/PhysRevLett.109.237008} {\bibfield
  {journal} {\bibinfo  {journal} {Phys. Rev. Lett.}\ }\textbf {\bibinfo
  {volume} {109}},\ \bibinfo {pages} {237008} (\bibinfo {year}
  {2012})}\BibitemShut {NoStop}%
\bibitem [{\citenamefont {Kuo}\ \emph {et~al.}(2015)\citenamefont {Kuo},
  \citenamefont {Tseng}, \citenamefont {Wang}, \citenamefont {Wang},
  \citenamefont {Chen}, \citenamefont {Wang}, \citenamefont {Lin},
  \citenamefont {Wu}, \citenamefont {Kuo},\ and\ \citenamefont
  {Lue}}]{PhysRevB.91.165141}%
  \BibitemOpen
  \bibfield  {author} {\bibinfo {author} {\bibfnamefont {C.~N.}\ \bibnamefont
  {Kuo}}, \bibinfo {author} {\bibfnamefont {C.~W.}\ \bibnamefont {Tseng}},
  \bibinfo {author} {\bibfnamefont {C.~M.}\ \bibnamefont {Wang}}, \bibinfo
  {author} {\bibfnamefont {C.~Y.}\ \bibnamefont {Wang}}, \bibinfo {author}
  {\bibfnamefont {Y.~R.}\ \bibnamefont {Chen}}, \bibinfo {author}
  {\bibfnamefont {L.~M.}\ \bibnamefont {Wang}}, \bibinfo {author}
  {\bibfnamefont {C.~F.}\ \bibnamefont {Lin}}, \bibinfo {author} {\bibfnamefont
  {K.~K.}\ \bibnamefont {Wu}}, \bibinfo {author} {\bibfnamefont {Y.~K.}\
  \bibnamefont {Kuo}}, \ and\ \bibinfo {author} {\bibfnamefont {C.~S.}\
  \bibnamefont {Lue}},\ }\href {\doibase 10.1103/PhysRevB.91.165141} {\bibfield
   {journal} {\bibinfo  {journal} {Phys. Rev. B}\ }\textbf {\bibinfo {volume}
  {91}},\ \bibinfo {pages} {165141} (\bibinfo {year} {2015})}\BibitemShut
  {NoStop}%
\bibitem [{\citenamefont {Homes}\ \emph {et~al.}(1993)\citenamefont {Homes},
  \citenamefont {Reedyk}, \citenamefont {Cradles},\ and\ \citenamefont
  {Timusk}}]{Homes:93}%
  \BibitemOpen
  \bibfield  {author} {\bibinfo {author} {\bibfnamefont {C.~C.}\ \bibnamefont
  {Homes}}, \bibinfo {author} {\bibfnamefont {M.}~\bibnamefont {Reedyk}},
  \bibinfo {author} {\bibfnamefont {D.~A.}\ \bibnamefont {Cradles}}, \ and\
  \bibinfo {author} {\bibfnamefont {T.}~\bibnamefont {Timusk}},\ }\href
  {\doibase 10.1364/AO.32.002976} {\bibfield  {journal} {\bibinfo  {journal}
  {Appl. Opt.}\ }\textbf {\bibinfo {volume} {32}},\ \bibinfo {pages} {2976}
  (\bibinfo {year} {1993})}\BibitemShut {NoStop}%
\bibitem [{\citenamefont {Tanner}(2015)}]{PhysRevB.91.035123}%
  \BibitemOpen
  \bibfield  {author} {\bibinfo {author} {\bibfnamefont {D.~B.}\ \bibnamefont
  {Tanner}},\ }\href {\doibase 10.1103/PhysRevB.91.035123} {\bibfield
  {journal} {\bibinfo  {journal} {Phys. Rev. B}\ }\textbf {\bibinfo {volume}
  {91}},\ \bibinfo {pages} {035123} (\bibinfo {year} {2015})}\BibitemShut
  {NoStop}%
\bibitem [{\citenamefont {Degiorgi}\ \emph {et~al.}(1996)\citenamefont
  {Degiorgi}, \citenamefont {Dressel}, \citenamefont {Schwartz}, \citenamefont
  {Alavi},\ and\ \citenamefont {Gr\"uner}}]{PhysRevLett.76.3838}%
  \BibitemOpen
  \bibfield  {author} {\bibinfo {author} {\bibfnamefont {L.}~\bibnamefont
  {Degiorgi}}, \bibinfo {author} {\bibfnamefont {M.}~\bibnamefont {Dressel}},
  \bibinfo {author} {\bibfnamefont {A.}~\bibnamefont {Schwartz}}, \bibinfo
  {author} {\bibfnamefont {B.}~\bibnamefont {Alavi}}, \ and\ \bibinfo {author}
  {\bibfnamefont {G.}~\bibnamefont {Gr\"uner}},\ }\href {\doibase
  10.1103/PhysRevLett.76.3838} {\bibfield  {journal} {\bibinfo  {journal}
  {Phys. Rev. Lett.}\ }\textbf {\bibinfo {volume} {76}},\ \bibinfo {pages}
  {3838} (\bibinfo {year} {1996})}\BibitemShut {NoStop}%
\bibitem [{\citenamefont {Hu}\ \emph {et~al.}(2008)\citenamefont {Hu},
  \citenamefont {Dong}, \citenamefont {Li}, \citenamefont {Li}, \citenamefont
  {Zheng}, \citenamefont {Chen}, \citenamefont {Luo},\ and\ \citenamefont
  {Wang}}]{PhysRevLett.101.257005}%
  \BibitemOpen
  \bibfield  {author} {\bibinfo {author} {\bibfnamefont {W.~Z.}\ \bibnamefont
  {Hu}}, \bibinfo {author} {\bibfnamefont {J.}~\bibnamefont {Dong}}, \bibinfo
  {author} {\bibfnamefont {G.}~\bibnamefont {Li}}, \bibinfo {author}
  {\bibfnamefont {Z.}~\bibnamefont {Li}}, \bibinfo {author} {\bibfnamefont
  {P.}~\bibnamefont {Zheng}}, \bibinfo {author} {\bibfnamefont {G.~F.}\
  \bibnamefont {Chen}}, \bibinfo {author} {\bibfnamefont {J.~L.}\ \bibnamefont
  {Luo}}, \ and\ \bibinfo {author} {\bibfnamefont {N.~L.}\ \bibnamefont
  {Wang}},\ }\href {\doibase 10.1103/PhysRevLett.101.257005} {\bibfield
  {journal} {\bibinfo  {journal} {Phys. Rev. Lett.}\ }\textbf {\bibinfo
  {volume} {101}},\ \bibinfo {pages} {257005} (\bibinfo {year}
  {2008})}\BibitemShut {NoStop}%
\bibitem [{\citenamefont {Fang}\ \emph {et~al.}(2013)\citenamefont {Fang},
  \citenamefont {Xu}, \citenamefont {Dong}, \citenamefont {Zheng},\ and\
  \citenamefont {Wang}}]{ISI:000314359400001}%
  \BibitemOpen
  \bibfield  {author} {\bibinfo {author} {\bibfnamefont {A.~F.}\ \bibnamefont
  {Fang}}, \bibinfo {author} {\bibfnamefont {G.}~\bibnamefont {Xu}}, \bibinfo
  {author} {\bibfnamefont {T.}~\bibnamefont {Dong}}, \bibinfo {author}
  {\bibfnamefont {P.}~\bibnamefont {Zheng}}, \ and\ \bibinfo {author}
  {\bibfnamefont {N.~L.}\ \bibnamefont {Wang}},\ }\href {\doibase
  {10.1038/srep01153}} {\bibfield  {journal} {\bibinfo  {journal} {{SCIENTIFIC
  REPORTS}}\ }\textbf {\bibinfo {volume} {{3}}} (\bibinfo {year} {{2013}}),\
  {10.1038/srep01153}}\BibitemShut {NoStop}%
\bibitem [{\citenamefont {Chen}\ \emph {et~al.}(2009)\citenamefont {Chen},
  \citenamefont {Xu}, \citenamefont {Hu}, \citenamefont {Zhang}, \citenamefont
  {Zheng}, \citenamefont {Chen}, \citenamefont {Luo}, \citenamefont {Fang},\
  and\ \citenamefont {Wang}}]{PhysRevB.80.094506}%
  \BibitemOpen
  \bibfield  {author} {\bibinfo {author} {\bibfnamefont {Z.~G.}\ \bibnamefont
  {Chen}}, \bibinfo {author} {\bibfnamefont {G.}~\bibnamefont {Xu}}, \bibinfo
  {author} {\bibfnamefont {W.~Z.}\ \bibnamefont {Hu}}, \bibinfo {author}
  {\bibfnamefont {X.~D.}\ \bibnamefont {Zhang}}, \bibinfo {author}
  {\bibfnamefont {P.}~\bibnamefont {Zheng}}, \bibinfo {author} {\bibfnamefont
  {G.~F.}\ \bibnamefont {Chen}}, \bibinfo {author} {\bibfnamefont {J.~L.}\
  \bibnamefont {Luo}}, \bibinfo {author} {\bibfnamefont {Z.}~\bibnamefont
  {Fang}}, \ and\ \bibinfo {author} {\bibfnamefont {N.~L.}\ \bibnamefont
  {Wang}},\ }\href {\doibase 10.1103/PhysRevB.80.094506} {\bibfield  {journal}
  {\bibinfo  {journal} {Phys. Rev. B}\ }\textbf {\bibinfo {volume} {80}},\
  \bibinfo {pages} {094506} (\bibinfo {year} {2009})}\BibitemShut {NoStop}%
\bibitem [{\citenamefont {Hu}\ \emph {et~al.}(2007)\citenamefont {Hu},
  \citenamefont {Li}, \citenamefont {Yan}, \citenamefont {Wen}, \citenamefont
  {Wu}, \citenamefont {Chen},\ and\ \citenamefont {Wang}}]{PhysRevB.76.045103}%
  \BibitemOpen
  \bibfield  {author} {\bibinfo {author} {\bibfnamefont {W.~Z.}\ \bibnamefont
  {Hu}}, \bibinfo {author} {\bibfnamefont {G.}~\bibnamefont {Li}}, \bibinfo
  {author} {\bibfnamefont {J.}~\bibnamefont {Yan}}, \bibinfo {author}
  {\bibfnamefont {H.~H.}\ \bibnamefont {Wen}}, \bibinfo {author} {\bibfnamefont
  {G.}~\bibnamefont {Wu}}, \bibinfo {author} {\bibfnamefont {X.}~\bibnamefont
  {Chen}}, \ and\ \bibinfo {author} {\bibfnamefont {N.~L.}\ \bibnamefont
  {Wang}},\ }\href {\doibase 10.1103/PhysRevB.76.045103} {\bibfield  {journal}
  {\bibinfo  {journal} {Phys. Rev. B}\ }\textbf {\bibinfo {volume} {76}},\
  \bibinfo {pages} {045103} (\bibinfo {year} {2007})}\BibitemShut {NoStop}%
\bibitem [{\citenamefont {Huang}\ \emph {et~al.}(2013)\citenamefont {Huang},
  \citenamefont {Wang}, \citenamefont {Wang}, \citenamefont {Shi},\ and\
  \citenamefont {Wang}}]{PhysRevB.87.100507}%
  \BibitemOpen
  \bibfield  {author} {\bibinfo {author} {\bibfnamefont {Y.}~\bibnamefont
  {Huang}}, \bibinfo {author} {\bibfnamefont {H.~P.}\ \bibnamefont {Wang}},
  \bibinfo {author} {\bibfnamefont {W.~D.}\ \bibnamefont {Wang}}, \bibinfo
  {author} {\bibfnamefont {Y.~G.}\ \bibnamefont {Shi}}, \ and\ \bibinfo
  {author} {\bibfnamefont {N.~L.}\ \bibnamefont {Wang}},\ }\href {\doibase
  10.1103/PhysRevB.87.100507} {\bibfield  {journal} {\bibinfo  {journal} {Phys.
  Rev. B}\ }\textbf {\bibinfo {volume} {87}},\ \bibinfo {pages} {100507}
  (\bibinfo {year} {2013})}\BibitemShut {NoStop}%
\bibitem [{\citenamefont {Huang}\ \emph {et~al.}(2014)\citenamefont {Huang},
  \citenamefont {Wang}, \citenamefont {Chen}, \citenamefont {Zhang},
  \citenamefont {Zheng}, \citenamefont {Shi},\ and\ \citenamefont
  {Wang}}]{PhysRevB.89.155120}%
  \BibitemOpen
  \bibfield  {author} {\bibinfo {author} {\bibfnamefont {Y.}~\bibnamefont
  {Huang}}, \bibinfo {author} {\bibfnamefont {H.~P.}\ \bibnamefont {Wang}},
  \bibinfo {author} {\bibfnamefont {R.~Y.}\ \bibnamefont {Chen}}, \bibinfo
  {author} {\bibfnamefont {X.}~\bibnamefont {Zhang}}, \bibinfo {author}
  {\bibfnamefont {P.}~\bibnamefont {Zheng}}, \bibinfo {author} {\bibfnamefont
  {Y.~G.}\ \bibnamefont {Shi}}, \ and\ \bibinfo {author} {\bibfnamefont
  {N.~L.}\ \bibnamefont {Wang}},\ }\href {\doibase 10.1103/PhysRevB.89.155120}
  {\bibfield  {journal} {\bibinfo  {journal} {Phys. Rev. B}\ }\textbf {\bibinfo
  {volume} {89}},\ \bibinfo {pages} {155120} (\bibinfo {year}
  {2014})}\BibitemShut {NoStop}%
\bibitem [{\citenamefont {H.~P.~Wang}\ and\ \citenamefont
  {Wang}(2014)}]{PhysRevB.90.144508}%
  \BibitemOpen
  \bibfield  {author} {\bibinfo {author} {\bibfnamefont {X.~B. W. Y. H. T. D.
  R. Y. C. G. H.~C.}\ \bibnamefont {H.~P.~Wang}, \bibfnamefont {Y.~L.~Sun}}\
  and\ \bibinfo {author} {\bibfnamefont {N.~L.}\ \bibnamefont {Wang}},\ }\href
  {\doibase 10.1103/PhysRevB.90.144508} {\bibfield  {journal} {\bibinfo
  {journal} {Phys. Rev. B}\ }\textbf {\bibinfo {volume} {90}},\ \bibinfo
  {pages} {144508} (\bibinfo {year} {2014})}\BibitemShut {NoStop}%
\bibitem [{\citenamefont {McMillan}(1976)}]{PhysRevB.16.643}%
  \BibitemOpen
  \bibfield  {author} {\bibinfo {author} {\bibfnamefont {W.~L.}\ \bibnamefont
  {McMillan}},\ }\href {\doibase 10.1103/PhysRevB.16.643} {\bibfield  {journal}
  {\bibinfo  {journal} {Phys. Rev. B}\ }\textbf {\bibinfo {volume} {16}},\
  \bibinfo {pages} {245119} (\bibinfo {year} {1976})}\BibitemShut {NoStop}%
\bibitem [{\citenamefont {Varma}\ and\ \citenamefont
  {Simons}(1983)}]{PhysRevLett.51.138}%
  \BibitemOpen
  \bibfield  {author} {\bibinfo {author} {\bibfnamefont {C.~M.}\ \bibnamefont
  {Varma}}\ and\ \bibinfo {author} {\bibfnamefont {A.~L.}\ \bibnamefont
  {Simons}},\ }\href {\doibase 10.1103/PhysRevLett.51.138} {\bibfield
  {journal} {\bibinfo  {journal} {Phys. Rev. Lett.}\ }\textbf {\bibinfo
  {volume} {51}},\ \bibinfo {pages} {138} (\bibinfo {year} {1983})}\BibitemShut
  {NoStop}%
\bibitem [{\citenamefont {X.Chen}()}]{PhysRevB.93.235121}%
  \BibitemOpen
  \bibfield  {author} {\bibinfo {author} {\bibnamefont {X.Chen}},\ }\href@noop
  {} {\bibfield  {journal} {\bibinfo  {journal} {Phys.rev.B}\ }\textbf
  {\bibinfo {volume} {93}},\ \bibinfo {pages} {235121}}\BibitemShut {NoStop}%
\end{thebibliography}%

\end{document}